\documentclass{ws-procs975x65}

\begin{document}

\title{Identifying Better Effective Higgsless Theories via $W_L$ $W_L$ Scattering}

\author{ Alexander S. Belyaev$^a$, R.\ Sekhar Chivukula$^b$, Neil D. Christensen$^b$,  \\
Hong-Jian He$^c$, Masafumi Kurachi$^d$\footnote{Speaker,\ \ 
Email: kurachi@tuhep.phys.tohoku.ac.jp}, 
Elizabeth H.\ Simmons$^b$, and Masaharu Tanabashi$^e$}

\address{$^a$ School of Physics \& Astronomy, University of Southampton,
Highfield, Southampton SO17 1BJ, UK\\
Particle Physics Department, Rutherford Appleton Laboratory, Chilton,
Didcot, Oxon OX11 0QX, UK\\
$^b$\ Department of Physics and Astronomy, Michigan State University, 
East Lansing, MI 48824, USA\\
$^c$\ Center for High Energy Physics, Tsinghua University, 
Beijing 100084, China\\
$^d$\ Department of Physics, Tohoku University, Sendai 980-8578, Japan\\
$^e$\ Department of Physics, Nagoya University, 
Nagoya 464-8602, Japan}

\begin{abstract}
The three site Higgsless model has been offered as a benchmark for studying the collider phenomenology 
of Higgsless models. In this talk, we present how well the three site Higgsless model performs as a general 
representative of Higgsless models, in describing $W_L W_L$ scattering, and which modifications 
can make it more representative.  We employ general sum rules relating the masses and couplings 
of the Kaluza-Klein (KK) modes of the gauge fields in continuum and deconstructed Higgsless 
models as a way to compare the different theories.  After comparing the three site Higgsless model to flat and 
warped continuum Higgsless models, we analyze an extensions of the three site Higgsless model, namely, 
the Hidden Local Symmetry (HLS) Higgsless model. We demonstrate that 
$W_LW_L$ scattering in the HLS Higgsless model can very closely approximate scattering in the 
continuum models, provided that the parameter `$a$'  is chosen to mimic $\rho$-meson dominance 
of $\pi\pi$ scattering in QCD. 
\end{abstract}
\keywords{Higgsless model; $W_L$$W_L$ scattering; Sum rule; Hidden Local Symmetry.}

\bodymatter

\section{Introduction}
Higgsless model \cite{Csaki:2003dt} is an attractive alternative to the Standard Model (SM) 
for describing the Electroweak symmetry breaking, 
in which the symmetry is broken by the boundary conditions of the five 
dimensional gauge theory. It turned out that dimensional deconstruction 
\cite{ArkaniHamed:2001ca} is quite useful to understand the important 
nature of the model, such as the delay of perturbative unitarity violation, 
boundary conditions, etc. (See Refs.\cite{SekharChivukula:2001hz, He:2004zr})
It was also extensively used to study the constraints from the electroweak 
precision measurements~\cite{Chivukula:2004pk}.

The three site Higgsless model \cite{SekharChivukula:2006cg} 
was proposed as an extremely deconstructed version of five dimensional Higgsless models, 
in which only one copy of weak gauge boson ($W'$, $Z'$) is introduced as a new resonance. 
Such a model contains sufficient complexity to incorporate interesting physics issues related to 
fermion masses and electroweak observables, yet remains simple enough that it could be 
encoded in a Matrix Element Generator program for use with Monte Carlo simulations. 
Such program was already done by several groups~\cite{He:2007ge}.
In this talk, we present how well the three site Higgsless model performs as a general 
representative of Higgsless models, in describing $W_L W_L$ scattering, and which modifications 
can make it more representative. After briefly reviewing the three site Higgsless model, 
we compare the three site Higgsless model to flat and warped continuum Higgsless models. 
Then, we analyze a Hidden Local Symmetry (HLS) generalization of the three site 
Higgsless model.\footnote{
This talk is based on the work done in Ref.~\cite{Belyaev:2009ve}}

\section{The three site Higgsless model}
\begin{figure}[t]
   \centering
   \includegraphics[width=0.6\textwidth]{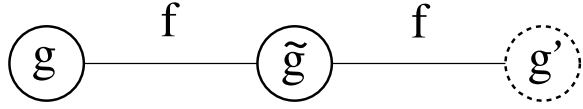} 
   \caption{Gauge sector of the three site Higgsless model. The solid circles represent
$SU(2)$ gauge groups, with coupling strengths $g_0$ and $g_1$, and the dashed circle is
a $U(1)$ gauge group with coupling $g_2$. }
   \label{fig:moose}
\end{figure}

In this section, we briefly review the three site Higgsless 
model\cite{SekharChivukula:2006cg}. 
The gauge sector of the three site Higgsless model 
is illustrated in Fig.~\ref{fig:moose} using ``Moose notation'' \cite{Georgi:1985hf}.  
The model  incorporates an
$SU(2) \times SU(2) \times U(1)$ gauge group (with couplings $g_0$, $g_1$ and $g_2$, 
respectively), and two  
nonlinear $(SU(2)\times SU(2))/SU(2)$ sigma models in which the global symmetry groups 
in adjacent sigma models are identified with the corresponding factors of the gauge group. 
The symmetry breaking between the middle $SU(2)$ and the $U(1)$ follows an 
$SU(2)_L \times SU(2)_R/SU(2)_V$ symmetry breaking pattern with the $U(1)$ 
embedded as the $T_3$-generator of $SU(2)_R$.    This extended electroweak gauge sector 
is  in the same class as models of extended  electroweak gauge symmetries 
\cite{Casalbuoni:1985kq}, which are considered as an application of the hidden 
local symmetry \cite{Bando:1984ej} to the electroweak sector.\footnote{
The new physics discussed in Ref.~\cite{SekharChivukula:2006cg} is related to the 
fermion sector. }
The decay constants, $f_1$ and $f_2$, of two nonlinear sigma models can be different 
in general, however, we take $f_1=f_2= f (= \sqrt{2}v)$ for simplicity.
Also, we work in the limit $x\equiv{g_0/g_1} \ll 1,\ y\equiv{g_2/g_1}\ll 1~$, 
in which case we expect a massless photon, light $W$ and $Z$ bosons, and
a heavy set of bosons $W'$ and $Z'$.  Numerically, then, $g_{0,2}$ are approximately
equal to the standard model $SU(2)_W$ and $U(1)_Y$ couplings, and we therefore
denote $g_0 \equiv g$ and $g_2 \equiv g'$, and define an angle $\theta$
such that 
$\frac{g'}{g} = \frac{\sin\theta}{\cos\theta} \equiv \frac{s}{c} \left(\equiv t \right)$.
In addition, we denote $g_1 \equiv \tilde{g}$.

\section{Comparison between continuum and the three site Higgsless models}

The three site Higgsless model can be viewed as an extremely deconstructed version 
of 5-dimensional $SU(2)\times SU(2)$ Higgsless model in which $SU(2)\times SU(2)$ 
symmetry is broken down to its diagonal $SU(2)$ by the boundary condition (BC) at one 
end of the extra dimension, while one of $SU(2)$ is broken down to its $U(1)$ subgroup 
by the BC at the other end. Thus, it is tempting to investigate how well the three site Higgsless 
model performs as a low energy effective theory of continuum Higgsless models.  
For this purpose, we consider $SU(2)\times SU(2)$ Higgsless 
model in the flat and the warped extra dimension as example models to compare.

The electroweak gauge sector of $SU(2)\times SU(2)$ Higgsless models 
in both the flat and the warped extra dimension can 
be characterized (with certain simplifications) by four free parameters. In flat case, those 
are $R$ (the size of the extra dimension),  $g_5$ (the bulk gauge coupling), $g_0$ and $g_Y$ 
(couplings for brane localized kinetic terms of $SU(2)\times U(1)$ gauge bosons). 
In warped case, those are $R'$ (the seize of extra dimension), $b$ (warp factor), $g_5$ and 
$g_Y$. (See Ref.~\cite{Chivukula:2005ji} for detailed descriptions of the models.) 
Since the gauge sector of the three site Higgsless model is also characterized by four 
free parameter (see, Fig.~\ref{fig:moose}), these three models can be compared by 
choosing four parameters so that they reproduce same values of four physical quantities. 
Three of four parameters should be chosen in a way that electroweak observables 
($e$, $M_Z$ and $M_W$, for example) are correctly reproduced. 
Then, there is still one free parameter to fix. Here, we use that parameter to fix 
the scale of KK mode (the mass of $W'$ boson, for example). Now, there are no free 
parameter left in each model, and any physical quantities other than those four quantities 
can be calculated as predictions in each models, which should be compared among 
three models to see how much the three site Higgsless model approximates continuum 
Higgsless models. 

In the present analysis, we focus on the values of triple gauge boson couplings each 
model predicts. As for triple gauge boson couplings which involve only SM gauge bosons, 
it can be shown that those are of the same form as in the SM at the leading order in the 
expansion of $M_W^2/M_{W'}^2$, and we require that the deviation from the SM value, which 
appears as next to leading order, should be within experimental bound (which put a 
lower bound on $M_{W'}$). 

In Table~\ref{tab:tbl1}, we listed the leading expressions 
of $Z'WW$ coupling and $ZW'W$ coupling in each models.
\begin{table}[t]
\tbl{Leading expressions 
of $Z'WW$ coupling and $ZW'W$ coupling in each models.}
{\begin{tabular}{|c|c|c|c|}
\hline
 & Three Site & 5D Flat & 5D Warped\\ 
\hline
$g_{Z'WW}$ 
& $-\frac{1}{2}\frac{e}{s}\left(\frac{M_W}{M_{W'}}\right)$ 
& $-\frac{4\sqrt{2}}{\pi^2}\frac{e}{s}\left(\frac{M_W}{M_{W'}}\right)$ 
& $-0.36\left(\frac{M_W}{M_{W'}}\right)$ 
\\ \hline
$g_{ZW'W}$ 
& $-\frac{1}{2}\frac{e}{sc}\left(\frac{M_W}{M_{W'}}\right)$ 
& $-\frac{4\sqrt{2}}{\pi^2}\frac{e}{sc}\left(\frac{M_W}{M_{W'}}\right)$ 
& $-0.36\frac{1}{c}\left(\frac{M_W}{M_{W'}}\right)$ 
\\ \hline
\end{tabular}}
\label{tab:tbl1}
\end{table}
Making the numerical approximations $\frac{4\sqrt{2}}{\pi^2}\frac{e}{s}\simeq 0.36$ 
and $\frac{4\sqrt{2}}{\pi^2}\frac{e}{s c}\simeq 0.41$, we find,
%
$
\frac{g_{Z'WW}|_{\rm warped-5d}}{g_{Z'WW}|_{\rm flat-5d}} \simeq 
\frac{g_{ZW'W}|_{\rm warped-5d}}{g_{ZW'W}|_{\rm flat-5d}} \simeq 1 .
\label{eq:similarities}
$
%
In other words, the values of $g_{Z'WW}$ and $g_{ZW'W}$ in these continuum models 
are essentially independent of the geometry of the extra dimension to leading order.
Then we can compare the couplings in continuum models to those in the three 
site Higgsless  model, assuming a common value for $M_W/M_{W'}$:
%
$
\frac{g_{Z'WW}|_{\rm three-site
}}{g_{Z'WW}|_{\rm flat-5d}} \simeq 
\frac{g_{ZW'W}|_{\rm three-site
}}{g_{ZW'W}|_{\rm flat-5d}} \simeq 
\frac{\pi^2}{8\sqrt{2}} \simeq 0.87 \,.
\label{eq:087}
$
%
The values of $g_{Z'WW}$ and $g_{ZW'W}$ in the three site
 Higgsless model are about 13\% smaller than those values in 5-dimensional $SU(2)\times SU(2)$ Higgsless models.  
Then, the question which naturally arises is --- ``Why do $g_{Z'WW}$ and $g_{ZW'W}$ take 
similar values in different models?" There are two keywords to be addressed to answer to this 
question: sum rules and the lowest KK mode dominance.

In any continuum five-dimensional gauge theory, the sum rules that guarantee the absence, 
respectively, of ${\cal{O}}(E^4)$ and ${\cal{O}}(E^2)$ growth in the amplitude for 
$W^+_L W^-_L \rightarrow W^+_L W^-_L$ elastic scattering have the following form 
\cite{Csaki:2003dt,SekharChivukula:2008mj},
\begin{equation}
 \sum_{i=1}^{\infty} g_{Z_{i}WW}^2 = g_{WWWW} - g_{ZWW}^2 - g_{\gamma WW}^2 \ ,
\label{eq:E4sum}
\end{equation}
\begin{equation}
 3 \sum_{i=1}^{\infty} g_{Z_{i}WW}^2M_{Z_{i}}^2
 = 4g_{WWWW}M_W^2  
 - 3g_{ZWW}^2M_Z^2 \ ,
\label{eq:E2sum}
\end{equation}
where $Z_i$ represents the $i$-th KK mode of the neutral gauge boson. ($Z_1$ is 
identified as $Z'$.)  We focuses on the degree to which the first KK mode saturates 
the sum on the LHS of the identities (\ref{eq:E4sum}) and (\ref{eq:E2sum}).   
Suppose that we form the ratio of the $n=1$ term in the sum on the LHS to the full 
combination of terms on the RHS, evaluated to leading order in $(M_W / M_{W'})^2$.  
The ratios derived from Eqs.~(\ref{eq:E4sum}) and (\ref{eq:E2sum}) are:
\begin{equation}
\frac{g_{Z'WW}^2}{g_{WWWW} - g_{ZWW}^2 - g_{\gamma WW}^2 } ~,
\label{eq:firstratio}
\end{equation}
\begin{equation}
\frac{3 g_{Z'WW}^2M_{Z'}^2}{ 4g_{WWWW}M_W^2  - 3g_{ZWW}^2M_Z^2}~.
\label{eq:secondratio}
\end{equation}
If the $n=1$ KK mode saturates the identity, then the related ratio will be 1.0;  
ratio values less than 1.0 reflect contributions from higher KK modes.  We see 
from Table~\ref{tab:tbl2} that each of these ratios is nearly 1.0 in both the 
$SU(2) \times SU(2)$ flat and warped Higgsless models, confirming that the 
first KK mode nearly saturates the sum rules in these continuum models.  
The similar behavior of the two extra dimensional models is consistent with our finding that 
the $g_{Z'WW}$ coupling is relatively independent of geometry.  

Because the first KK mode nearly saturates the identities (\ref{eq:E4sum}) and 
(\ref{eq:E2sum}) in these continuum models, the ratios (\ref{eq:firstratio}) and 
(\ref{eq:secondratio}) should be useful for drawing comparisons with the three 
site Higgsless model, which only possesses a single KK gauge mode.  As shown in the 3rd 
column of Table~\ref{tab:tbl2}, the first ratio has the value one in the three site 
Higgsless model, 
meaning that the identity  (\ref{eq:E4sum}) is still satisfied.  The ratio related to 
identity (\ref{eq:E2sum}), however, has the value 3/4 for the three site Higgsless model, 
meaning that the second identity is not satisfied; the longitudinal gauge boson 
scattering amplitude continues to grow as $E^2$ due to the underlying 
non-renormalizable interactions in the three site Higgsless model. Since the value of the 
denominator in Eq.~(\ref{eq:secondratio}) has not changed appreciably, this indicates a difference between 
the values of the $g_{Z'WW}$ couplings in the continuum and three site Higgsless models. 
This is the reason why $g_{Z'WW}$ in the three site Higgsless model is 13\% 
(or 25\% in $g_{Z'WW}^2$) smaller than the value of $g_{Z'WW}$ in 
continuum models as we have shown in the previous section.

\begin{table}[t]
\tbl{Ratios relevant to evaluating the degree of cancellation of growth in the 
$W_L W_L$ scattering amplitude from the lowest lying $KK$ resonance at 
order $E^4$ (top row, from Eq. (\ref{eq:firstratio})), and at order $E^2$ (second 
row, from Eq. (\ref{eq:secondratio})). A value close to one indicates a high 
degree of cancellation from the lowest lying resonance. Shown in successive 
columns for the $SU(2) \times SU(2)$ flat and warped continuum models 
discussed in the text, and the three site Higgsless model.}
{\begin{tabular}{|c|c|c|c|}
\hline
&\  5d $2\times 2$ Flat\  &\  5d $2\times 2$ Warped\  &\  Three-site\ 
 \\ \hline
$ \frac{g_{Z'WW}^2}{g_{WWWW} - g_{ZWW}^2 - g_{\gamma WW}^2 }$
& 
$\frac{960}{\pi^6} \simeq 0.999$
&
$0.992$
&
$1$ 
\\ \hline
$ \frac{3 g_{Z'WW}^2M_{Z'}^2
 }{ 4g_{WWWW}M_W^2  
 - 3g_{ZWW}^2M_Z^2}$
&
$\frac{96}{\pi^4} \simeq 0.986$
&
$0.986$
&
3/4
\\  \hline
\end{tabular}}
\label{tab:tbl2}
\end{table}

\section{Hidden Local Symmetry generalization of the three site Higgsless models}
In this section, we consider an Hidden Local Symmetry (HLS) generalization of the 
three site Higgsless model, in which an extra parameter $a$ is introduced and the 
Lagrangian takes the following form\cite{Bando:1984ej}:
\begin{eqnarray}
{\cal L}^{\rm HLS} &=& -\,\frac{v^2}{4} {\rm Tr} \left[(D^\mu \Sigma^\dagger_1) \Sigma_1
- (D^\mu \Sigma_2) \Sigma^\dagger_2\right]^2 -a\,\frac{v^2}{4} {\rm Tr}\,
\left[(D^\mu \Sigma^\dagger_1) \Sigma_1 + (D^\mu \Sigma_2) \Sigma^\dagger_2\right]^2, \nonumber \\
&=& \frac{v^2}{4}(1+a) {\rm Tr} \left[ (D_\mu \Sigma_1)^\dagger(D^\mu \Sigma_1)  
+   (D_\mu \Sigma_2)^\dagger(D^\mu \Sigma_2)  \right] \nonumber \\
&&\ \  + \frac{v^2}{2}(1-a) {\rm Tr} \left[ (D_\mu \Sigma_1)^\dagger 
\Sigma_1 (D_\mu \Sigma_2) \Sigma_2^\dagger  \right] ,\nonumber \\
&=& \frac{a v^2 }{2} {\rm Tr} \left[ (D_\mu \Sigma_1)^\dagger(D^\mu \Sigma_1)  
+   (D_\mu \Sigma_2)^\dagger(D^\mu \Sigma_2) \right]  \nonumber \\
&&\ \ + \frac{v^2}{4}(1-a) \,  {\rm Tr} \left[ (D_\mu (\Sigma_1\Sigma_2))^\dagger  
D^\mu (\Sigma_1\Sigma_2)   \right]\,. 
\label{eq:HLS}
\end{eqnarray}
Note that taking $a=1$ reproduce the Lagrangian of the three site Higgsless model. 
Fermion sector of the model is also discussed in detail in Ref.~\cite{Belyaev:2009ve}, in which 
we showed that the model can accommodate the ideal fermion delocalization which is needed 
for the consistency with the precision EW experiments.

It is straightforward to calculate triple gauge boson couplings of this model, from which we 
can evaluate the values in Eqs.~(\ref{eq:firstratio}) and (\ref{eq:secondratio}):
\begin{equation}
\frac{g_{Z'WW}^2}{g_{WWWW} - g_{ZWW}^2 - g_{\gamma WW}^2 }  = 1, \ \ \ \ 
\frac{3 g_{Z'WW}^2M_{Z'}^2}{ 4g_{WWWW}M_W^2  - 3g_{ZWW}^2M_Z^2} 
= \frac{3}{4}\, a \,.
\end{equation}
By comparing this result with the ones shown in Table~\ref{tab:tbl2}, 
we see that the HLS Higgsless model can very closely approximate scattering in the 
continuum models if we take $a=\frac{4}{3}$.

It is interesting to note that $a=\frac{4}{3}$ is the choice in which the four-point 
Nambu-Goldstone (NG) boson coupling, $g_{\pi\pi\pi\pi} = 1-\frac{3}{4}a$, vanishes in 
this model.\cite{Harada:2003jx}
Considering that, in the language of dimensional 
deconstruction\cite{ArkaniHamed:2001ca}, 
NG bosons are identified as a fifth-component of the gauge boson field ($A_5$) in a 
discretized five-dimensional gauge theory, and also considering the fact that there is 
no four-point $A_5$ coupling in five dimensional gauge theories, it is natural that the 
parameter choice in which the four-point NG boson coupling vanishes well approximates 
the continuum models.

\begin{figure}[t]
  \begin{center}
    \includegraphics[width=0.6\textwidth]{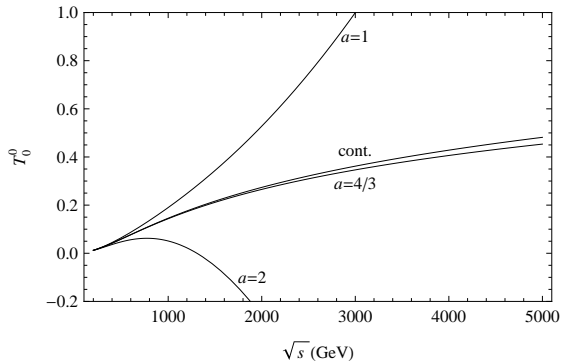}
  \end{center}
\caption{Behavior of the partial wave amplitude $T_0^0$ for NG boson scattering 
in the triangular moose model with 
various $a$. The values $v=250$ GeV, $M_1=500$ GeV are assumed. The curve labeled ``cont." shows 
$T_0^0$ in the continuum flat $SU(2)\times SU(2)$ model for $M_1 = 500$ GeV.}
\label{fig:t00}
\end{figure}
Fig.~\ref{fig:t00} shows the partial wave  amplitude $T_0^0$ for NG boson scattering 
in the global (which means we set $g = g' = 0$ for simplicity) continuum flat 
$SU(2)\times SU(2)$ model (with the mass of the first KK mode, $M_1$, taken to be  
$M_1=500$ GeV) compared with $T_0^0$ in the HLS Higgsless model 
for several values of the parameter $a$. 
The result in the global
 three site Higgsless model is shown by the curve labeled $a=1$; 
the value $a=2$ is motivated by the phenomenological 
KSRF relation \cite{Kawarabayashi:1966kd}.
This plot ties our results together quite neatly: while the curves with three different 
values of $a$ all give a reasonable description of $T_0^0$ at very low energies, 
the best approximation to the continuum 
behavior of $T_0^0$ over a wide range of energies is given by the HLS Higgsless  
model curve with $a=4/3$.   At low energies, the fact that the three-site and  the HLS 
Higgsless models both prevent $E^4$ growth of the amplitude suffices; but at higher 
energies, the fact that the HLS Higgsless model with $a = 4/3$ has 
$g_{\pi\pi\pi\pi} = 0$ and enables it to cut off the $E^2$ growth of the amplitude 
as well, as consistent with the behavior in the continuum model.

\section{Conclusions}

We have considered how well the three site Higgsless model performs as a general representative 
of Higgsless models, and have studied HLS generalization have the potential to improve 
upon its performance.  Our comparisons have employed sum rules 
relating the masses and couplings of the gauge field KK modes.   
We find that the tendency of the sum rules to be saturated by contributions from the 
lowest-lying KK resonances suggests a way to quantify the extent to which a 
highly-deconstructed theory like the three site Higgsless model can accurately describe the 
low-energy physics.  We demonstrated that $W_LW_L$ scattering in the HLS Higgsless 
model can very closely approximate scattering in the continuum models, provided that 
the HLS parameter $a$ is chosen appropriately. 
This observation confirms that the collider phenomenology 
studied, such as in Ref.~\cite{He:2007ge} , are applicable not only to the three site Higgsless 
model, but also to extra-dimensional Higgsless models.

\bibliographystyle{ws-procs975x65}
\bibliography{ws-pro-sample}

\begin{thebibliography}{9}

\bibitem{Csaki:2003dt}
  C.~Csaki, C.~Grojean, H.~Murayama, L.~Pilo and J.~Terning,
  Phys.\ Rev.\  D {\bf 69}, 055006 (2004); 
  C.~Csaki, C.~Grojean, L.~Pilo and J.~Terning,
  Phys.\ Rev.\ Lett.\  {\bf 92}, 101802 (2004).

\bibitem{ArkaniHamed:2001ca}
  N.~Arkani-Hamed, A.~G.~Cohen and H.~Georgi,
  Phys.\ Rev.\ Lett.\  {\bf 86}, 4757 (2001); 
  C.~T.~Hill, S.~Pokorski and J.~Wang,
  Phys.\ Rev.\  D {\bf 64}, 105005 (2001).
  
\bibitem{SekharChivukula:2001hz}
  R.~S.~Chivukula, D.~A.~Dicus and H.~J.~He,
  Phys.\ Lett.\  B {\bf 525}, 175 (2002);
  R.~S.~Chivukula and H.~J.~He,
  Phys.\ Lett.\  B {\bf 532}, 121 (2002).
  
\bibitem{He:2004zr}
  H.~J.~He,
  Int.\ J.\ Mod.\ Phys.\  A {\bf 20} (2005) 3362.
    
\bibitem{Chivukula:2004pk}
  R.~S.~Chivukula, E.~H.~Simmons, H.~J.~He, M.~Kurachi and M.~Tanabashi,
  Phys.\ Rev.\  D {\bf 70}, 075008 (2004); 
  Phys.\ Lett.\  B {\bf 603}, 210 (2004); 
  Phys.\ Rev.\  D {\bf 71}, 035007 (2005); 
  Phys.\ Rev.\  D {\bf 71}, 115001 (2005); 
  Phys.\ Rev.\  D {\bf 72}, 015008 (2005); 
  Phys.\ Rev.\  D {\bf 72}, 095013 (2005).
    
\bibitem{SekharChivukula:2006cg}
  R.~Sekhar Chivukula, B.~Coleppa, S.~Di Chiara, E.~H.~Simmons, H.~J.~He, M.~Kurachi and M.~Tanabashi,
  Phys.\ Rev.\  D {\bf 74}, 075011 (2006).
  
\bibitem{He:2007ge}
  H.~J.~He {\it et al.},
  Phys.\ Rev.\  D {\bf 78}, 031701 (2008); 
  J.~G.~Bian {\it et al.},
  Nucl.\ Phys.\  B {\bf 819}, 201 (2009).
  
\bibitem{Belyaev:2009ve}
  A.~S.~Belyaev, R.~Sekhar Chivukula, N.~D.~Christensen, H.~J.~He, M.~Kurachi, E.~H.~Simmons and M.~Tanabashi,
  Phys.\ Rev.\  D {\bf 80}, 055022 (2009).

\bibitem{Georgi:1985hf}
  H.~Georgi,
  Nucl.\ Phys.\  B {\bf 266}, 274 (1986).

\bibitem{Casalbuoni:1985kq}
  R.~Casalbuoni, S.~De Curtis, D.~Dominici and R.~Gatto,
  Phys.\ Lett.\  B {\bf 155}, 95 (1985); 
  R.~Casalbuoni, A.~Deandrea, S.~De Curtis, D.~Dominici, R.~Gatto and M.~Grazzini,
  Phys.\ Rev.\  D {\bf 53}, 5201 (1996).
  
\bibitem{Bando:1984ej}
  M.~Bando, T.~Kugo, S.~Uehara, K.~Yamawaki and T.~Yanagida,
  Phys.\ Rev.\ Lett.\  {\bf 54}, 1215 (1985); 
  M.~Bando, T.~Kugo and K.~Yamawaki,
  Nucl.\ Phys.\  B {\bf 259}, 493 (1985); 
  M.~Bando, T.~Fujiwara and K.~Yamawaki,
  Prog.\ Theor.\ Phys.\  {\bf 79}, 1140 (1988); 
  M.~Bando, T.~Kugo and K.~Yamawaki,
  Phys.\ Rept.\  {\bf 164}, 217 (1988).

\bibitem{Chivukula:2005ji}
  R.~S.~Chivukula, E.~H.~Simmons, H.~J.~He, M.~Kurachi and M.~Tanabashi,
  Phys.\ Rev.\  D {\bf 72}, 075012 (2005).
  
\bibitem{SekharChivukula:2008mj}
  R.~S.~Chivukula, H.~J.~He, M.~Kurachi, E.~H.~Simmons and M.~Tanabashi,
  Phys.\ Rev.\  D {\bf 78}, 095003 (2008).


\bibitem{Harada:2003jx}
See section 3 in  M.~Harada and K.~Yamawaki,
 Phys.\ Rept.\  {\bf 381}, 1 (2003); 
The relation between $a=4/3$ and $\rho$ meson dominance was also 
pointed out in M.~Harada, S.~Matsuzaki and K.~Yamawaki,
 Phys.\ Rev.\  D {\bf 74}, 076004 (2006).
 
\bibitem{Kawarabayashi:1966kd}
  K.~Kawarabayashi and M.~Suzuki,
  Phys.\ Rev.\ Lett.\  {\bf 16}, 255 (1966); 
  Riazuddin and Fayyazuddin,
  Phys.\ Rev.\  {\bf 147}, 1071 (1966).

  
\end{thebibliography}


\end{document}